# "Bose minus Einstein" and other examples of misattribution

V. Hushwater [a]

*70 St. Botolph St, apt. 401, Boston, MA 02116*

Following the paper of J.D. Jackson on a false attribution of discoveries, equations, and methods [Am. J. Phys. **76**, 704-719 (2008)] I consider examples, in which some of independent discoverers are forgotten, and in other cases discoveries and effects are attributed to people who did not discover them at all. I also discuss in brief why that happens and how physics community can avoid this.

## I. INTRODUCTION

In an interesting paper[1] J.D. Jackson gave a few examples on a false attribution of discoveries, equations, and methods to people who had not done them first. Following that paper I consider below examples, in which some of independent discoverers are forgotten, and in other cases discoveries and effects are attributed to people who did not discover them at all! I conclude by discussing in brief why that happens and how physics community can avoid this.

## II. EXAMPLES WHEN SOME OF INDEPENDENT DISCOVERERS ARE FORGOTTEN

### A. Landau levels

The energy levels of a nonrelativistic charged particle in homogeneous magnetic field were determined by L. Landau in 1930. Now they are called *Landau levels*. But he has predecessors: in 1928 V. Fock solved a problem of an oscillator in external magnetic field and I. Rabi solved a problem of a Dirac electron in external magnetic field. Also, J. Frenkel' and M. Bronstein solved in 1930 the Schrödinger eq. of a nonrelativistic charged particle in homogeneous magnetic field independently from L. Landau, see a paper of V.L. Pokrovsky[2] and references therein. Pokrovsky thinks that only Landau's name became attached to such levels because none of other papers established a connection of the quantization with diamagnetism and oscillations of the diamagnetic susceptibility.

### B. Landau-Zener transitions.

Let a Hamiltonian of a system $H$ depends on a parameter $R$. The discreet energy levels $E_n(R)$ and vectors of the state (eigenfunctions) $\Psi_n(R)$ corresponding to them continuously depend on that parameter. If $R$ varies with time $t$, the energy is not conserved and also depends on time. However, if the parameter $R$ varies slowly, the vector of state is equal to $\Psi_n(R(t))$ in the leading approximation, i.e. the system follows the continuous change of $R$ with time without making a transition to other states. This is the so-called adiabatic approximation. It is not valid if some of the levels cross. Near the crossing point of the levels there are intensive transitions between the corresponding states. In the case that only two levels cross and other levels are far away, such transitions are usually called *Landau-*

*Zener transitions*. The problem of such transitions was formulated and solved in 1932 independently by L. Landau, and C. Zener. But, not only by them, in the same year the solution of the problem was published also by E.G.G. Stuckelberg, and E. Majorana, see Ref. 2.

## III. EXAMPLES WHEN DISCOVERIES AND EFFECTS ATTRIBUTED TO PEOPLE WHO DID NOT DISCOVER THEM AT ALL

### A. Dynamic(al) or Nonstationary Casimir Effect

In 1948 H. B. G. Casimir predicted that there is an attractive force between two uncharged, parallel, perfectly conducted plates in vacuum,[3, 4] which was later measured experimentally. The force is a result of a spectral redistribution of normal modes of quantum vacuum fluctuations of the electromagnetic field between plates in comparison to the free space.[5] As a consequence of such redistribution the vacuum radiation pressure on a plate from between the plates is smaller than the radiation pressure on a plate from outside. This phenomenon is rightfully called *the Casimir Effect*.

If separation of plates changes very slowly normal modes of quantum fluctuations of the electromagnetic field will remain in the ground state, only their frequencies will be shifted to values corresponding to changing boundary conditions.[5] However, if separation of plates changes fast normal modes of quantum fluctuations of the electromagnetic field will not remain in the ground state making transitions to excited states, i.e. generating real photons. This will also happen for any fast changes of the boundary conditions. A qualitative explanation of such phenomena is the parametric amplification of quantum fluctuations of the electromagnetic field in systems with time-dependent parameters.[6] Due to back-reaction of emitted photons on plates they experience a dissipative force. All these phenomena are often but unjustifiably called *the Dynamic(al)* or *Nonstationary Casimir Effect*. The creation of photons by moving mirrors was first predicted by G. T. Moore[7] in 1970 for a one-dimensional cavity. S. A. Fulling and P. C. W. Davis[8] showed in 1976 that there is a dissipative force even for a single mirror in 1+1 space-time. After these two papers many more physically realistic treatments were published, see references in reviews[6, 9] and in a monograph[10] (Ch. 4). That is why the authors of a review[11] suggest calling the creation of photons by moving mirrors as *the Moore effect*.

As is written in Ref. 6, "The reference to vacuum fluctuations explains the appearance of Casimir's name (by analogy with the famous static Casimir effect, which is also considered frequently as a manifestation of quantum vacuum fluctuations …), although Casimir himself did not write anything on this subject." I think that the reference to vacuum fluctuations is not enough to justify such an appearance. There are many phenomena caused by vacuum fluctuations and by modification of them due to boundaries, e. g. in the field of research called "Cavity QED", which takes its origin from an abstract of E. M. Purcell[12] published in 1946, i.e. before the paper of Casimir.[3] For short history see Ref. 13.

For completeness I should add that one can question whether papers[7, 8] should be considered as first papers on the phenomena under consideration? A similar phenomenon of particle creation in the expanding universe has been studied already by E. Schrödinger[14] in 1939; see also Ch. 3 and 5 in Ref. 10.

## B. Bose-Einstein (Bose *minus* Einstein): From Einstein condensation to Bose condensation and beyond…

In 1925 A. Einstein, based on his theory of the ideal gas of (what is now called) bosons,[15] predicted[16] that below a certain temperature a finite fraction of gas atoms will occupy the zero momentum level (ground state). Einstein called such a phenomenon *condensation* by analogy with the ordinary condensation of a gas in coordinate space. Correspondingly, he called the fraction of gas on the zero momentum level *condensate*. The reality of that phenomenon was uncertain for many years due to objections to Einstein's mathematical treatment and arguments that all real gases become liquids or solids at the temperatures in question. Later, however, Einstein's conclusion was supported by more elaborate mathematics,[17–19] and from 1995 the phenomenon became a hot topic both experimentally and theoretically (for a weakly interacting gas). The history of that development is discussed in short in Refs. 20 - 22. There should not be any question about who made the discovery. In early monographs and textbooks, which discuss quantum statistics: A. Sommerfeld[23] (published in German in1930s), D. ter Haar[24] (written in 1954), and C. Kittel,[25] they use the historically correct term, *Einstein condensation*. In his next textbook[26] Kittel keeps this attribution of the discovery to Einstein.

However F. London[17] called the phenomenon, *Bose-Einstein condensation*, in spite of the fact that he stated in the text that discovery was made by only Einstein. Also, in the Russian translation[27] of Kittel's book,[26] they replaced his original *Einstein condensation* not just by *Bose-Einstein condensation* (once), but, more often, by *Bose condensation* or *bose-condensation* . <u>This really appears to be "**Bose minus Einstein**" – S. P. Kapitsa, the editor of translation of the book in the Soviet Union, deliberately "**subtracted**" Einstein's name from the title of the phenomenon and replaced it by Bose!</u> However, not only he but C. Kittel himself faltered under peer pressure. In his second edition of the book written together with H. H. Krommer,[28] they use the term *Bose-Einstein condensation* (while still referring to the temperature of the condensation as *Einstein condensation temperature*).

Thus for a long time the phenomenon and the corresponding state are unjustifiably called *Bose-Einstein* (or just *Bose*) *condensation/condensate* instead of *Einstein condensation/condensate* (*of bosons*). Still some historically conscientious authors rebel, e.g.: "This phenomenon is called **Einstein condensation** (sometimes, illegitimately, Bose-Einstein condensation)." [29] Many people, who know history and even refer to Einstein papers on the subject, use the term *Bose-Einstein condensation* just by analogy with *Bose-Einstein statistics*. But we have to separate the discovery of the statistics, to which both Bose and Einstein contributed and the discovery of the *condensation*, to which Bose did not and could not contribute. (He has dealt only with photons - massless particles, which number is not conserved and, therefore, they cannot form a *condensate*.) Others, especially younger ones, who do not know history, erroneously believe that,[30] "The phenomenon of Bose-Einstein condensation [was] initially predicted by Bose [1] and Einstein [2, 3] in 1924…"

But a sin against Einstein does not stop here. Physicists like short names and therefore erroneously use *Bose statistics* instead of *Bose-Einstein statistics*, *Bose gas* instead of *bosonic gas* or *gas of bosons*. The usage of *Bose gas* instead of *bosonic gas* disregards the fact that Bose, as I mentioned above, created and applied his statistics only for photons, and never even worked on atomic gases. It was Einstein who courageously generalized the Bose method for massive particles and developed the statistical mechanics of an ideal gas.[15] But Bose's name is so short and catchy that many cannot resist the urge to attach it to more and more terms, e. g., *Bose-condensed gases* (or *fluids*), *Bose superfluids*, *Bose order parameter*… Probably farther than most went the author of a textbook:[31] "The difference between the Planck distribution and Maxwell-Boltzman distribution can be attributed to [a statistical] attraction, as Einstein showed, in what has come to be known by the historical but clumsy name of 'theory of A and B coefficients.' Consider a two-level atom in the wall of a black-body cavity… 'Bose enhancement' refers to the fact that [the] rate of this process is enhanced by the presence of photons of the same frequency." So the author suggests replacing "the historical but clumsy name" by a catchy but unhistorical one. First of all, as the author acknowledged, the theory was created by Einstein[32] (in 1916), and is usually referred as *Einstein's A and B coefficients*.[33] If the author thinks that he needs a term, which includes a word *enhancement*, why in such a case not to use a historically correct and not clumsy term, *Einstein enhancement*? Bose does not need such an undeserved gift – he made his important contribution in discovering the statistics of photons and that statistics rightly carries his name.

## IV. CONCLUDING REMARKS

I agree with J. D. Jackson, that in most cases misattribution occurs "because the community was not diligent in searching the prior literature before attaching a name to the discovery or relation or effect."[1] Of course the great volume of new knowledge and accelerated rate of acquiring it can be partially blamed for this, as well as the intensity of competition with its current motto, "publish or perish." In addition, as noted by D. Politzer,[34] "The use of history in science education may be a contributing factor to why this is so and how it works. As teachers of the next generation of scientists, we always seek to compress and simplify all the developments that have come before. We want to bring our students as quickly as possible to the frontier of current understanding. From this perspective, the actual history, which involves many variants and many missteps, is only a hindrance. And the neat, linear progress, as outlined by the sequence of gleaming gems… is a useful fiction. But a fiction it is. The truth is often far more complicated." So it appears that interest in history of physics – real history – is little or absent among many physicists. This is sad since looking back can help in seeing what is ahead. Knowing the history of ideas can help in better understanding nuances of established concepts, in understanding trends, in not repeating earlier mistakes, in training imagination etc. In order to help in awakening interest in the history of physics, in addition to papers and books devoted to it, authors of physics monographs, reviews and resource letters should discuss, let briefly, a development of ideas and experiments.

However, the last two examples show that it is not only the absence of knowledge of actual history that leads to misattribution. Some people know and even refer to real authors, but for unknown reasons attach someone else's name to the discovery or relation or effect. Maybe the reason is not sociological – peer pressure, but psychological – people just do not care…

But the right attribution is first of all a question of fairness and honesty. Rephrasing Newton, we stand on shoulders (and heads) of earlier generations of researchers and thinkers. We should not forget their contributions. Development of science is not just a race for acquiring more knowledge and understanding – it is a part of human culture. As I.I. Rabi wrote:[35] "I propose that science be taught, at whatever level, from the lowest to the highest, in the humanistic way. It should be taught with a certain historical understanding, with a certain philosophical understanding, with a social understanding and a human understanding, in the sense of the biography, the nature of the people who make this construction, the triumphs, the trials, the tribulations." Of course such an education is very difficult to achieve in our time, but we should strive for this.

Finally, often ideas and understanding are a result of wandering in "blind alleys" and are clarified and grow from one author to another, see e.g. the history of asymptotic freedom[34]. In these cases it is impossible to figure out, who contributed to a discovery more even if people who made a final step first are known. They demonstrate validity of Berry's law:[36] "Nothing is ever discovered for the first time." In such cases we should not attach any person's name to the discovery or relation or effect. The chosen term should express their essence and appropriate references should be provided.

**Acknowledgments**

I thank Jim Swank for useful comments and remarks.

[a] Electronic mail: quantcosmos@yahoo.com
[1] J.D. Jackson, "Examples of the zeroth theorem of the history of science," Am. J. Phys. **76**, 704-719 (2008).
[2] V.L. Pokrovsky, "Landau and modern physics," Physics-Uspekhi **52** (11), 1169-1176 (2009).
[3] H. B. G. Casimir, "On the attraction between two perfectly conducting plates," Proc. K. Ned. Akad. Wet. **51**, 793-795 (1948).
[4] P. W. Milonni, "The Quantum Vacuum: An Introduction to Quantum Electrodynamics," (Academic, Boston, 1994).
[5] V. Hushwater, "Repulsive Casimir force as a result of vacuum radiation pressure," Am. J. Phys. **65**, 381 (1997); "Radiation Pressure Approach to the Repulsive Casimir Force," Com. Atom. Mol. Phys., Com. Mod. Phys. **1**, Part D, 329-335 (2000).
[6] V V Dodonov, "Current status of the dynamical Casimir effect," Phys. Scr. **82**, 038105, 1-10 (2010); http://arxiv.org/abs/1004.3301v2 [quant-ph].
[7] G. T. Moore, "Quantum Theory of the Electromagnetic Field in a Variable-Length One-dimensional Cavity," J. Math. Phys. **11**, 2679-2691 (1970).
[8] S. A. Fulling, P. C. W. Davis, "Radiation from a Moving Mirror in Two Dimensional Space-Time: Conformal Anomaly," Proc. R. Soc. London, A **348**, 393-414 (1976).
[9] M. Kardar, R. Golestanian, "The "friction" of vacuum, and other fluctuation-induced forces," Rev. Mod. Phys. **71** (4) 1233-1245 (1999).
[10] Birrell and C. W. Davis, *Quantum fields in curved space* (Cambridge U. P., Cambridge, 1984).
[11] L. C. B. Crispino, A. Higuchi, G. E. A. Matsas, "The Unruh effect and its applications," Rev. Mod. Phys. **80**, 787-838 (2008).
[12] E. M. Purcell, "Spontaneous Emission Probabilities at Radio Frequencies," Phys. Rev. **69**, 681 (1946).
[13] S. Haroche and J.-M. Raimond, *Exploring the Quantum: Atoms, Cavities, and Photons* (Oxford U. P., Oxford, 2006) pp. 232-250.
[14] E. Schrödinger, "The Proper Vibrations of the Expanding Universe," Physica **VI** (9), 899-912 (1939).
[15] A. Einstein, "Quantentheorie des einatomigen idealen Gases," Sitzungsber. Preuss. Akad. Wiss., Phys.-math. Kl, 261-267 (1924).
[16] A. Einstein, "Quantentheorie des einatomigen idealen Gases. Zweite Abhandlung," Sitzungsber. Preuss.

Akad. Wiss., Phys.-math. Kl**,** 3-14 (1925).
[17] F. London, "On the Bose-Einstein Condensation," Phys. Rev., **54,** 947- 954 (1938).
[18] B. Kahn B. and G.E. Uhlenbeck, "On the Theory of Condensation," Physica, no 5, 399-416 (1938).
[19] G.E. Uhlenbeck, "Some Reminiscences about Einstein's Visits to Leiden," in *Some Strangeness in the Proportion*, edit. H. Woolf (Addison-Wesley, London, 1980) pp. 524-525.
[20] A. Pais, *Subtle is the Lord…* (Oxford U. P., N. Y., 1982) pp. 432-434.
[21] A. Griffin, "A Brief History of Our Understanding of BEC: From Bose to Beliaev," in *Bose-Einstein Condensation in Atomic Gases*, ed. M. Inguscio, S. Stringari and C. Wieman (IOS Press, Amsterdam, Netherlands, 1999) pp. 1-13; cond-mat/9901123.
[22] J. Stachel, "Einstein and Bose," Talk at Bose Centenary Meeting, Calcutta 1994, in J. Stachel, *Einstein from 'B' to 'Z'* (Birkhauser, Boston, 2002) pp. 519-538.
[23] A. Sommerfeld, *Thermodynamics and Statistical Mechanics* (Academic, New York, 1956) p. 273.
[24] D. ter Haar, *Elements of Statistical Mechanics* (Rinehart & Company, New York, 1958) pp. 89, 209.
[25] C. Kittel, *Elementary Statistical Physics* (John Wiley & Sons, Ney York, 1958) p. 100.
[26] C. Kittel, *Thermal Physics* (Wiley, New York, 1969) pp. 273, 275, 278, 281.
[27] C. Kittel, *Staisticheskaya Termodinamika* (*Statistical Thermodynamics*) (I. L., Moscow, 1977) [in Russian].
[28] C. Kittel, H. Krommer, *Thermal Physics* (W. H. Freeman, San Francisco, 1980) pp. 202, 205, 218.
[29] L. Peliti, *Statistical Mechanics in a Nutshell* (Princeton U. P., Princeton and Oxford, 2011) p. 113.
[30] P. G. Kevrekidis et al, "Basic Mean-Field Theory for Bose-Einstein Condensates," in *Emergent Nonlinear Phenomena in Bose-Einstein Condensates: Theory and Experiment*, edit. P. G. Kevrekidis et al (Springer, N. Y., 2008) pp. 3-21.
[31] K. Huang, *Introduction to Statistical Physics* (CRC Press, Boca Raton, FL, 2010) pp. 239, 240.
[32] A. Einstein, "Strahlungs-Emission und-Absorption nach der Quantentheorie," Verh. Deutsch. Phys. Ges. **18**, 318-323 (1916).
[33] R. Loudon, *The Quantum Theory of Light* (Oxford U. P., Oxford, 1997).
[34] D. Politzer, "Nobel Lecture: The dilemma of attribution," Rev. Mod. Phys., **77**, 851-856 (2004).
[35] G. Holton, "I.I. Rabi as Educator and Science Warrior," Phys. Today, Sep., 37-42 (1999).
[36] M. V. Berry, "Three laws of discovery," in Questions at <www.phy.bris.ac.uk/people/berry-mv/>.